\documentclass[journal,twoside]{IEEEtran}
\usepackage[T1]{fontenc}
\usepackage{amsmath,amssymb,amsthm}
\usepackage{xcolor}
\usepackage{cite}
\usepackage[hidelinks]{hyperref}
\usepackage{microtype}
\usepackage{array,booktabs,tabularx}
\newtheorem{theorem}{Theorem}
\newtheorem{corollary}[theorem]{Corollary}
\newtheorem{lemma}[theorem]{Lemma}
\newtheorem{proposition}[theorem]{Proposition}
\newtheorem{example}[theorem]{Example}
\theoremstyle{definition}
\newtheorem{definition}{Definition}
\theoremstyle{plain}
\DeclareMathOperator{\Jac}{Jac}
\DeclareMathOperator{\Pic}{Pic}
\DeclareMathOperator{\supp}{supp}
\newcommand{\F}{\mathbb F}
\begin{document}
\color{black}
\title{MDS Algebraic Geometry Codes\\of Square-Root Length on Hyperelliptic Curves}
\author{{Chuangqiang Hu} and {Guorui Ma}%
\thanks{\raggedright Chuangqiang Hu is with Sun Yat-sen University, Guangzhou 510275, China (e-mail: huchq5@mail.sysu.edu.cn).}%
\thanks{Guorui Ma is with the Shanghai Institute for Mathematics and Interdisciplinary Sciences (SIMIS) and the Research Institute of Intelligent Complex Systems, Fudan University, Shanghai 200433, China (e-mail: mgr18@tsinghua.org.cn). Guorui Ma is the corresponding author.}}
\maketitle

\begin{abstract}

For every fixed genus $g\geq2$ and every $0<\rho<1$, we construct maximum distance separable (MDS) algebraic geometry codes over $\F_q$ for all sufficiently large odd squares $q$. Their parameters are $\bigl[n,\lfloor\rho n\rfloor,n-\lfloor\rho n\rfloor+1\bigr]_q$ with $n=\sqrt q+1+O_g(q^{1/4})$. The codes use full Riemann--Roch spaces on genus-$g$ hyperelliptic curves. They are not monomially equivalent to generalized Reed--Solomon (GRS) codes when $q$ is sufficiently large. An infinite subfamily has exact length $n=\sqrt q+2gq^{1/4}+1$.
\end{abstract}


\section{Introduction}
Reed--Solomon codes are obtained by evaluating polynomials of bounded degree over a finite field \cite{ReedSolomon1960}. Their minimum distance meets the Singleton bound. Goppa extended this construction to algebraic curves by replacing polynomials with Riemann--Roch spaces \cite{Goppa1981}. This gave algebraic geometry codes, whose parameters reflect both the genus and the rational points of the curve. A standard account appears in \cite[Ch.~2]{Stichtenoth2009}.

Write $\F_s$ for a finite field with $s$ elements. A linear $[n,k,d]_s$ code is a $k$-dimensional subspace of $\F_s^n$ with minimum Hamming distance $d$. Its rate is $k/n$. The Singleton bound is $d\leq n-k+1$, and a code attaining equality is called maximum distance separable (MDS). On a curve of genus $g$, the usual designed-distance bound for an algebraic geometry code can be smaller than $n-k+1$ by $g$. To obtain an MDS code, one must rule out functions that vanish at too many evaluation points. Thus a large number of rational points alone does not ensure the MDS property.

\begingroup
MDS codes from curves of positive genus have been studied through divisor classes, point configurations, and code equivalence. Munuera proved the main conjecture for elliptic codes and for genus-two codes when the field size exceeds $83$ \cite[Thms.~1--2]{Munuera1992}. His Proposition~1 reduces the length problem to a condition on representatives of classes of sums of distinct evaluation points. He also studied length bounds for elliptic MDS codes \cite{Munuera1993}. The hyperelliptic case of the main conjecture, under hypotheses on the genus and field size, was treated independently by Chen \cite{ChenMainConj1994} and de Boer \cite{deBoer1996}. Walker developed an approach through projective embeddings and proved the elliptic case by this method \cite{Walker1996}. These upper-bound results explain the role of divisor classes. The present paper addresses existence with a specified length and dimension range.

In arbitrary genus, Chen gave an MDS criterion and a counting construction \cite[Thms.~4.2--4.3]{Chen2024}. For a one-point divisor of degree $m\geq4g$, the dimension is $k=m-g+1$. The sufficient condition for length $n>m$ is
\[
 m\binom{n}{m-g}<N,
\]
where $N\leq q+1+2g\sqrt q$ is the number of rational points in the projective image. For fixed $g$ and $k/n\to\rho\in(0,1)$, put
\[
 H(\rho)=-\rho\log\rho-(1-\rho)\log(1-\rho)>0.
\]
Here $\log$ denotes the natural logarithm. Stirling's formula gives
\[
 \log\!\left(m\binom{n}{m-g}\right)=nH(\rho)+o(n).
\]
Thus this sufficient condition can certify only $n=O(\log q)$ at a fixed limiting rate. This observation concerns that counting condition; it is not an upper bound for all codes in \cite{Chen2024}.

Marques, da Silva, and Tafazolian study complementary pairs on hyperelliptic curves \cite{Marques2026}. Their Proposition~6 characterizes the MDS property of their length-$2t$, dimension-$t$ pairs by requiring reduced Mumford representatives of degree $g$ for classes associated with every $t$-subset of the evaluation set. The present construction proves simultaneous avoidance for all such subset classes by placing them in a subgroup of the Jacobian. It imposes no complementarity or LCD condition. 

Zhu, Chen, Peng, Zhang, and Zhao study MDS and NMDS constructions from maximal curves and genus-two curves with split Jacobian \cite{Zhu2026}. Their results impose determinant, evaluation-point, elliptic-cover, and subset-sum conditions, including $k\equiv2\pmod3$ in one specialization. {\color{black}For the Beelen--Montanucci branch, write $q=a^{2h}$, where $a$ is a prime power and $h\geq3$ is odd, and put
\[
 s=\frac{a^{h-1}-1}{a^2-1}.
\]
Theorem~1 of \cite{Zhu2026} uses a curve of genus
\[
 \frac{(a-1)(a^{h+1}+a^h-a^2)}{2}
\]
and a one-point divisor giving dimension $s+3$. This genus varies with $a$ and $h$; for example, $h=3$ gives $s=1$ and dimension $4$, while the genus grows with $a$. Their genus-two branch instead fixes the genus at two and assumes a split Jacobian, with sufficient conditions on the evaluation set.} We do not claim to improve every length in \cite{Zhu2026}. Our contribution is a uniform existence theorem in every fixed genus, over all sufficiently large odd square fields, with $g+1\leq k\leq n-g$, full Riemann--Roch spaces, and eventual non-GRS codes. Table~\ref{tab:comparison} records the scope of these comparisons.
\endgroup

The question considered here is the following. For every fixed $g\geq2$, can one obtain algebraic geometry codes over $\F_q$ with length asymptotic to $\sqrt q$ and with rate bounded away from zero and one? We require the full Riemann--Roch space of a divisor. This requirement excludes constructions obtained by selecting a proper subspace of that space. We allow general defining divisors, rather than requiring all poles to lie at one point.

{{\color{black}Throughout, curves are smooth, projective, and geometrically irreducible; an affine equation denotes its smooth projective model. The construction assumes $g\geq2$.

The evaluation points are defined over $\F_r$, and the defining divisor is defined over $\F_{r^2}$. Classes of sums of evaluation points lie in the subfield Jacobian. Corollary~\ref{cor:coset-mds} reduces the MDS condition to finding a coset that avoids effective divisor classes of degree $g-1$. The main geometric step is Proposition~\ref{prop:coset}. Its proof uses the intersection formula in Lemma~\ref{lem:intersection} twice, through Lemma~\ref{lem:fixed-point-descent}, to reduce the degree of a fixed point of a translated Frobenius involution by two and ultimately exclude all such effective classes.}}

\begin{theorem}
\label{thm:main}
Let $r$ be an odd prime power, let $g\geq2$, and assume $r\geq(2g+2)^2$. Let $X/\F_r$ be a hyperelliptic curve of genus $g$ with an odd-degree affine model
\begin{equation}
 \begin{gathered}
 y^2=f(x),\qquad \deg f=2g+1,\\
 f\in\F_r[x]\ \text{square-free}.
 \end{gathered}
 \label{eq:model}
\end{equation}
Set $q=r^2$, let $n=|X(\F_r)|$, and let $D$ be the sum of the points in $X(\F_r)$. For every integer
\begin{equation}
 {g+1\leq k\leq n-g},
 \label{eq:dimension}
\end{equation}
there is an effective $\F_q$-rational divisor $G$ of degree $k+g-1$, with support disjoint from $D$, such that $C_{\mathcal L}(D,G)$ is an $[n,k,n-k+1]_q$ MDS code.
\end{theorem}

{Proposition~\ref{prop:non-grs} below gives the additional non-equivalence assertion: if $\min\{k,n-k\}\geq3g+1$, then the code is not monomially equivalent to a generalized Reed--Solomon code.}

The theorem applies to each curve with the stated model. Existence of such curves in every genus gives the following consequence. In the asymptotic statements, the genus is fixed and $q$ tends to infinity through odd squares.

{
\begin{corollary}
\label{cor:rate}
Fix $g\geq2$ and $0<\rho<1$. For every sufficiently large odd square $q$, there are genus-$g$ MDS algebraic geometry codes, not monomially equivalent to generalized Reed--Solomon codes, with parameters
\begin{equation}
 \left[n,\lfloor\rho n\rfloor,n-\lfloor\rho n\rfloor+1\right]_q,
 \label{eq:rho-parameters}
\end{equation}
where $n$ satisfies
\begin{equation}
 \sqrt q+1-2gq^{1/4}\leq n\leq\sqrt q+1+2gq^{1/4}.
 \label{eq:length}
\end{equation}
Writing $k=\lfloor\rho n\rfloor$, we have
\begin{equation}
 \frac{k}{n}=\rho+O(n^{-1})\longrightarrow\rho,\qquad
 n=\sqrt q+1+O_g\!\left(q^{1/4}\right).
 \label{eq:rates}
\end{equation}
\end{corollary}
}

The next consequence uses a fixed equation and gives an exact length.

{
\begin{corollary}
\label{cor:exact}
Fix $g\geq2$, $0<\rho<1$, and a prime $p\equiv-1\pmod{2(2g+1)}$. For every sufficiently large positive odd integer $m$, put $q=p^{4m}$. On the curve
\begin{equation}
 X:y^2=1-x^{2g+1}
 \label{eq:maximal}
\end{equation}
there are MDS codes, not monomially equivalent to generalized Reed--Solomon codes, with parameters \eqref{eq:rho-parameters} and
\begin{equation}
 n=\sqrt q+2gq^{1/4}+1.
 \label{eq:exact}
\end{equation}
\end{corollary}
}

{A recent preprint gives $n\leq q/2+o(q)$ for fixed genus when both the dimension and the codimension grow faster than $\log q$ \cite[Thm.~1]{ChenHuMa2026}. That growth condition includes the families in Corollary~\ref{cor:rate}. There remains a gap in order between that upper bound and the present existence lower bound. No upper bound of order $\sqrt q$, or optimal leading constant, is proved here.}

The proof has three parts. Section~\ref{sec:prelim} expresses the MDS condition in the Jacobian and constructs effective representatives with the required support. Section~\ref{sec:geometry} proves the hyperelliptic intersection formula and the coset statement. Section~\ref{sec:proofs} finds the required points, proves the main results, and checks the examples. Section~\ref{sec:future} states the remaining questions.

\begin{table*}[t]

\caption{Scope of the closest MDS constructions.}
\label{tab:comparison}
\centering\footnotesize
\setlength{\tabcolsep}{5pt}
\renewcommand{\arraystretch}{1.18}
\begin{tabularx}{\textwidth}{@{}>{\raggedright\arraybackslash}p{0.17\textwidth}XX@{}}
\toprule
Result & Curves and defining divisors & Length and dimension guarantee\\
\midrule
Chen \cite[Thm.~4.3]{Chen2024}
& Arbitrary genus; one-point divisor.
& $m\binom{n}{m-g}<N$; at fixed rate, only $n=O(\log q)$.\\
Zhu et al. \cite[Thm.~1, Cor.~3]{Zhu2026}
& Beelen--Montanucci curves; one-point divisors.
& Determinant/evaluation conditions; dimension $s+3$; genus varies with $a,h$.\\
Zhu et al. \cite[Thm.~4, Cors.~5--6, 11]{Zhu2026}
& Genus two with split Jacobian.
& Elliptic-cover and subset conditions; $k\equiv2\pmod3$ in Corollary~11; length depends on the point set.\\
Marques et al. \cite[Prop.~6]{Marques2026}
& Hyperelliptic curves; complementary pairs.
& Length $2t$, dimension $t$; reduced-degree MDS condition for every $t$-subset.\\
Theorem~\ref{thm:main}, Corollary~\ref{cor:rate}
& Hyperelliptic curves satisfying \eqref{eq:model}; general effective divisors.
& $g+1\leq k\leq n-g$, $n=\sqrt q+1+O_g(q^{1/4})$.\\
\bottomrule
\end{tabularx}
\end{table*}

\section{Preliminaries}
\label{sec:prelim}
\subsection{Divisors and Evaluation Codes}
Let $K$ be a finite field and let $X/K$ be a curve of genus $g$. A divisor on $X$ is a finite sum $H=\sum_T m_TT$, where $T$ runs through closed points and $m_T\in\mathbb Z$. Its degree is $\sum_T m_T[K(T):K]$, where $K(T)$ is the residue field at $T$. It is effective, written $H\geq0$, if every $m_T$ is nonnegative. Its support $\supp(H)$ is the set of points with nonzero coefficient. Equivalently, a $K$-rational divisor is a divisor on the geometric curve invariant under the Galois group of an algebraic closure $\overline K$ over $K$. It need not be a sum of points in $X(K)$.

Write $K(X)$ for the function field of $X$. For $h\in K(X)^\times$, its principal divisor is $(h)=\sum_T v_T(h)T$, where $v_T$ is the normalized order at $T$. Two $K$-rational divisors are linearly equivalent, written $H\sim H'$, if $H-H'=(h)$ for such an $h$. Set
\begin{equation}
 \begin{split}
 \mathcal L(H)&=\{h\in K(X)^\times:(h)+H\geq0\}\cup\{0\},\\
 \ell(H)&=\dim_K\mathcal L(H).
 \end{split}
 \label{eq:RRspace}
\end{equation}
The complete linear system $|H|$ consists of effective divisors linearly equivalent to $H$. If $K_X$ is a canonical divisor, Riemann--Roch gives
\begin{equation}
 \ell(H)=\deg H+1-g+\ell(K_X-H),\qquad \deg K_X=2g-2.
 \label{eq:RR}
\end{equation}
In particular, $\ell(H)=0$ if $\deg H<0$, and $\ell(H)=\deg H+1-g$ if $\deg H>2g-2$; see \cite[Sec.~1.5]{Stichtenoth2009}.

{
\begin{definition}
\label{def:ag-code}
Fix distinct $P_1,\ldots,P_n\in X(K)$, and put $D=P_1+\cdots+P_n$. If $\supp(G)\cap\supp(D)=\varnothing$, every function in $\mathcal L(G)$ is regular at every $P_i$. Define the evaluation map and the algebraic geometry code by
\begin{equation}
 \begin{split}
 \operatorname{ev}_D:\mathcal L(G)&\longrightarrow K^n,\\
 h&\longmapsto(h(P_1),\ldots,h(P_n)),\\
 C_{\mathcal L}(D,G)&=\operatorname{ev}_D(\mathcal L(G)).
 \end{split}
 \label{eq:ag-code}
\end{equation}
\end{definition}
}
Its minimum distance is the smallest number of nonzero coordinates of a nonzero codeword.

\begingroup
The next standard criterion identifies the precise vanishing condition we must prove. The use of these vanishing spaces is classical; see \cite[Sec.~I]{Munuera1992} and \cite[Thm.~4.2]{Chen2024}. We retain the proof for general $G$ to make the subsequent sufficient condition self-contained.
\endgroup

\begin{lemma}
\label{lem:ag-code-criterion}
Suppose $\deg G<n$, and set $k=\ell(G)\geq1$. Then $C_{\mathcal L}(D,G)$ has dimension $k$ and minimum distance at least $n-\deg G$. It is MDS if and only if
\begin{equation}
 \ell(G-E)=0
 \label{eq:mds-criterion}
\end{equation}
for every sum $E$ of $k$ distinct points among $P_1,\ldots,P_n$.
\end{lemma}

\begin{proof}
The kernel of \eqref{eq:ag-code} is $\mathcal L(G-D)$. Its degree is negative, so this space is zero. Thus evaluation is injective and its image has dimension $k$.

Let $h\in\mathcal L(G)$ be nonzero, and let $Z$ be the sum of the evaluation points at which $h$ vanishes. The support condition on $G$ implies $h\in\mathcal L(G-Z)$. Consequently, $\deg Z\leq\deg G$, since a divisor of negative degree has no nonzero section. The corresponding codeword has weight $n-\deg Z\geq n-\deg G$.

For a sum $E$ of $k$ distinct evaluation points, the kernel of evaluation on just those coordinates is $\mathcal L(G-E)$. If \eqref{eq:mds-criterion} holds for all such $E$, no nonzero codeword has $k$ zero coordinates. Its weight is therefore at least $n-k+1$. The Singleton bound gives equality. Conversely, a nonzero element of $\mathcal L(G-E)$ evaluates to a nonzero codeword, by injectivity, with at least $k$ zero coordinates. Its weight is at most $n-k$, so the code is not MDS.
\end{proof}

\subsection{Effective Divisor Classes in the Jacobian}
We now express \eqref{eq:mds-criterion} in a form that uses the subfield of definition of the evaluation points. Fix $P_0\in X(K)$. Let $J=\Jac(X)$ be the Jacobian. Write $[H]$ for the linear-equivalence class of a divisor $H$. The group $J(K)$ identifies with the degree-zero divisor classes represented over $K$. More generally, $\Pic^d(X)(K)$ denotes the degree-$d$ line-bundle classes, and translation by $dP_0$ identifies it with $J(K)$. The existence of $P_0$ ensures that these rational classes have line-bundle, and hence divisor, representatives over $K$; see \cite[Part~III, Sec.~1]{Milne2008}.

For an integer $d\geq0$, let $X^{(d)}$ be the $d$-th symmetric power of $X$; let $X^{(0)}$ be a point defined over $K$. Its $K$-points are effective $K$-rational divisors of degree $d$. Consider the Abel--Jacobi map
\begin{equation}
 \alpha_d:X^{(d)}\longrightarrow J,\qquad A\longmapsto[A-dP_0].
 \label{eq:abel}
\end{equation}
{
\begin{definition}
\label{def:Wd}
Let $W_d$ be the closed image of $\alpha_d$. Its set of $K$-rational points is
\begin{equation}
 \begin{split}
 W_d(K)=\{[A-dP_0]:\;& A\geq0,\ \deg A=d,\\
 & A\text{ defined over }K\}.
 \end{split}
 \label{eq:Wd}
\end{equation}
\end{definition}
The standard construction and these maps are described in \cite[Part~III, Secs.~3--5]{Milne2008}.
}
{Concretely, we must rule out the possibility that a class has an effective representative over $\overline K$ but not over $K$. The class is represented by a line bundle $L$ of degree $d$ defined over $K$. If $L$ has a nonzero section over $\overline K$, then its section space is already nonzero over $K$, because the dimension of this space does not change under field extension. Thus a nonzero section exists over $K$, and its zero divisor is an effective $K$-rational representative of the class.}

{On geometric points, the fibers of \eqref{eq:abel} are complete linear systems: for an effective divisor $A$ of degree $d$,
\begin{equation}
 \alpha_d^{-1}\bigl([A-dP_0]\bigr)=|A|.
 \label{eq:abel-fiber}
\end{equation}
Indeed, $\alpha_d(B)=\alpha_d(A)$ if and only if $B\sim A$.}

In particular, $W_0(K)=\{0\}$, and adding $P_0$ to an effective divisor gives
\begin{equation}
 W_{d-1}(K)\subseteq W_d(K)\qquad(d\geq1).
 \label{eq:nesting}
\end{equation}
The effective classes in $\Pic^{g-1}(X)$ form the theta divisor. Under the translation above, its inverse image in $J$ is $W_{g-1}$ \cite[Part~III, Secs.~5--6]{Milne2008}. We use this translated theta divisor throughout.

The following lemma isolates the implication from the geometric construction to the coding problem. This is the only place where the classes of sums of evaluation points are used.

\begin{lemma}
\label{lem:coset-criterion}
Assume $X$, $P_0$, and $P_1,\ldots,P_n$ are defined over $\F_r\subseteq K$. Let $b\in J(K)$ satisfy
\begin{equation}
 (b+J(\F_r))\cap W_{g-1}(K)=\varnothing.
 \label{eq:coset}
\end{equation}
For an integer $1\leq k\leq n$, let $G$ be a $K$-rational divisor with
\begin{equation}
 \deg G=k+g-1,\qquad [G-(k+g-1)P_0]=b.
 \label{eq:G-class}
\end{equation}
Then $\ell(G-E)=0$ for every sum $E$ of $k$ distinct evaluation points.
\end{lemma}

\begin{proof}
For such an $E$, put $e=[E-kP_0]\in J(\F_r)$. If $\ell(G-E)>0$, choose $0\ne h\in\mathcal L(G-E)$. Then $A=(h)+G-E$ is an effective $K$-rational divisor of degree $g-1$. Equations \eqref{eq:G-class} and \eqref{eq:Wd} give
\[
 [A-(g-1)P_0]=b-e\in(b+J(\F_r))\cap W_{g-1}(K),
\]
contradicting \eqref{eq:coset}.
\end{proof}

\subsection{Effective Representatives with Disjoint Support}
The coset argument produces a divisor class. Evaluation requires an actual divisor whose support avoids $D$. The next elementary argument provides an effective representative without assuming that it is supported at rational points.

\begin{lemma}
\label{lem:effective-representative}
Let $|K|=q$ and assume $n<q$. For every $b\in J(K)$ and $k\geq g+1$, there is an effective $K$-rational divisor $G$ satisfying \eqref{eq:G-class}, with $\supp(G)\cap\supp(D)=\varnothing$ and $\ell(G)=k$.
\end{lemma}

\begin{proof}
Set $a=k+g-1\geq2g$. Represent $b$ by a $K$-rational divisor $B$ of degree zero and put $G_0=B+aP_0$. Then $\deg G_0=a$ and $[G_0-aP_0]=b$. Since $a>2g-2$, Riemann--Roch gives $\ell(G_0)=a+1-g=k$. For each evaluation point $R$, the divisor $G_0-R$ has degree at least $2g-1$, so the same formula gives
\begin{equation}
 \ell(G_0-R)=k-1.
 \label{eq:hyperplanes}
\end{equation}
Thus each $\mathcal L(G_0-R)$ is a proper subspace with $q^{k-1}$ elements. Their union has at most $nq^{k-1}<q^k$ elements. Choose $h\in\mathcal L(G_0)$ outside this union. In particular $h\ne0$.

Set $G=G_0+(h)$. It is effective because $h\in\mathcal L(G_0)$, and it has degree $a$ and the required class. At an evaluation point $R$, let $c_R$ be the coefficient of $G_0$. Membership in $\mathcal L(G_0)$ gives $v_R(h)+c_R\geq0$. Nonmembership in $\mathcal L(G_0-R)$ gives $v_R(h)+c_R<1$. This integer is therefore zero, which is exactly the coefficient of $R$ in $G$. Hence the supports are disjoint. Finally, linear equivalence gives $\ell(G)=\ell(G_0)=k$.
\end{proof}
\begingroup
Combining these two lemmas gives the following coset sufficient condition. It reduces the coding problem to geometric avoidance; the existence of the required coset is proved in Proposition~\ref{prop:coset}.
\endgroup
{
\begin{corollary}
\label{cor:coset-mds}
Assume that $X$, $P_0$, and $P_1,\ldots,P_n$ are defined over $\F_r\subseteq K$, where $|K|=q$ and $n<q$. Let $b\in J(K)$ satisfy \eqref{eq:coset}. For every integer $g+1\leq k\leq n-g$, there is an effective $K$-rational divisor $G$ of degree $k+g-1$, with support disjoint from $D$, such that $C_{\mathcal L}(D,G)$ is an $[n,k,n-k+1]_K$ MDS code.
\end{corollary}

\begin{proof}
Lemma~\ref{lem:effective-representative} gives an effective $K$-rational divisor $G$ satisfying \eqref{eq:G-class}, with support disjoint from $D$ and $\ell(G)=k$. Lemma~\ref{lem:coset-criterion} gives $\ell(G-E)=0$ for every sum $E$ of $k$ distinct evaluation points. Since $\deg G=k+g-1<n$, Lemma~\ref{lem:ag-code-criterion} proves the MDS property.
\end{proof}
}

\section{Hyperelliptic Curves and the Coset Construction}
\label{sec:geometry}
{The goal of this section is to construct a Jacobian coset disjoint from the translated theta divisor. More precisely, we seek $b\in J(\F_{r^2})$ satisfying $(b+J(\F_r))\cap W_{g-1}(\F_{r^2})=\varnothing$. We first use the hyperelliptic involution to prove an intersection formula for $W_d(K)$, and then combine it with Frobenius and fixed-point descent to obtain the required coset.}
\subsection{Divisors of Small Degree}
{Let $g\geq2$, let $K$ be a finite field of odd characteristic, and let $X/K$ be the smooth projective model of $y^2=f(x)$, where $f\in K[x]$ is square-free and $\deg f=2g+1$.} The degree-two map $x:X\to\mathbb P^1$ to the projective line is branched at the $2g+1$ roots of $f$ and at infinity. The Hurwitz formula gives genus $g$. The point above infinity is unique, is $K$-rational, and will be denoted by $P_0$. The hyperelliptic involution is $\iota(x,y)=(x,-y)$, extended to $X$. For every finite geometric point $P$, the divisor of $x-x(P)$ gives
\begin{equation}
 P+\iota(P)\sim2P_0.
 \label{eq:fiber}
\end{equation}
At a branch point the zero has multiplicity two. The same equivalence is immediate for $P=P_0$. These are the fibers of the map $x$.

Apply $\iota$ to each point of a divisor. For an effective $K$-rational divisor $A$ of degree $d$, relation \eqref{eq:fiber} gives $A+\iota(A)\sim2dP_0$. The equivalence is defined over $K$, either by multiplying the functions for the points in Galois orbits or by pulling back a divisor on $\mathbb P^1$. Consequently,
\begin{equation}
 -[A-dP_0]=[\iota(A)-dP_0],\qquad -W_d(K)=W_d(K).
 \label{eq:symmetry}
\end{equation}

{To control intersections of translates of $W_d$, we first prove a structural fact about Riemann--Roch spaces of divisors of degree at most $g$. Its consequence concerning hyperelliptic fibers is used below.}

\begin{lemma}
\label{lem:low-degree}
{Let $F$ be an effective $K$-rational divisor with $\deg F\leq g$. If $\ell(F)\geq2$, then $\mathcal L(F)\subseteq K(x)$.}
\end{lemma}

\begin{proof}
Since $F$ is effective, constants belong to $\mathcal L(F)$. Choose a nonconstant $h\in\mathcal L(F)$, and let $e=\deg(h)_\infty$, where $(h)_\infty$ is its pole divisor. The associated map $h:X\to\mathbb P^1$ has degree $e$, and $1\leq e\leq\deg F\leq g$.

Suppose $h\notin K(x)$. The extension $K(X)/K(x)$ has degree two, so $K(x,h)=K(X)$. Thus $(x,h)$ is birational onto an integral curve $C$ in $\mathbb P^1\times\mathbb P^1$. Its two projection degrees are $2$ and $e$. A curve with these projection degrees has arithmetic genus
\[
 p_a(C)=(2-1)(e-1)=e-1.
\]
For completeness, put $S=\mathbb P^1\times\mathbb P^1$ and write $\mathcal O_S(a,b)$ for the tensor product of the pullbacks of the degree-$a$ and degree-$b$ line bundles from the two factors. Write $\chi$ for the alternating sum of cohomology dimensions. The exact sequence
\[
 0\longrightarrow\mathcal O_S(-e,-2)\longrightarrow
 \mathcal O_S\longrightarrow\mathcal O_C\longrightarrow0
\]
and $\chi(\mathcal O_S(a,b))=(a+1)(b+1)$ give this formula (with the factors interchanged if necessary). The normalization of $C$ is $X$, and normalization can only decrease arithmetic genus. Hence $g\leq e-1$, contradicting $e\leq g$.

It follows that $h\in K(x)$, as required.
\end{proof}

{
\begin{corollary}
\label{cor:low-degree-fiber}
Let $F$ be an effective $K$-rational divisor with $\deg F\leq g$. If $\ell(F)\geq2$, then $F-2P_0$ is linearly equivalent over $K$ to an effective divisor.
\end{corollary}

\begin{proof}
Choose a nonconstant $h\in\mathcal L(F)$. Lemma~\ref{lem:low-degree} gives $h\in K(x)$. Let $T$ be its pole divisor on $\mathbb P^1$, and set $s=\deg T\geq1$. Then $(h)_\infty=x^*T\sim2sP_0$. Since $(h)_\infty\leq F$, we have
\[
 F-2P_0\sim F-(h)_\infty+2(s-1)P_0\geq0,
\]
as required. All the divisors and equivalences used here are defined over $K$.
\end{proof}
}

\subsection{Frobenius and the Required Coset}
Assume now that $X$ is defined over $\F_r$ and $K=\F_{r^2}$. Let $\sigma$ be the $r$-power Frobenius endomorphism, acting on $X$ and on $J$. On $K$-points it has square equal to the identity, and its fixed subgroup on $J(K)$ is $J(\F_r)$. Since $P_0$ is rational over $\F_r$, we have $\sigma(W_d(K))=W_d(K)$.

{
Since $\sigma$ and $\iota$ commute, $\iota\sigma$ is an involution on $X(K)$.
{
\begin{definition}
\label{def:iota-sigma-fixed}
A point $P\in X(K)$ is called an \emph{$\iota\sigma$-fixed point} if $\iota(\sigma(P))=P$, or equivalently $\sigma(P)=\iota(P)$.
\end{definition}
}
Let $P,Q$ be two such points with finite distinct $x$-coordinates, so that
\begin{equation}
 \sigma(P)=\iota(P),\qquad \sigma(Q)=\iota(Q).
 \label{eq:points}
\end{equation}
Put
\begin{equation}
 u=[P-P_0],\qquad v=[Q-P_0],\qquad
 \delta=u+v.
 \label{eq:point-classes}
\end{equation}
We first record the relations between these divisor classes and then construct the required coset as a set of fixed points.

\begin{lemma}
\label{lem:point-classes}
{With this notation,} $u\ne v$, $\delta\ne0$, and
\begin{equation}
 \sigma(u)=-u,\qquad \sigma(v)=-v,\qquad
 \sigma(\delta)=-\delta.
 \label{eq:delta}
\end{equation}
\end{lemma}

\begin{proof}
Since $P_0$ is defined over $\F_r$, equations \eqref{eq:fiber} and \eqref{eq:points} give
\[
 \sigma(u)=[\iota(P)-P_0]=-[P-P_0]=-u.
\]
The same computation gives $\sigma(v)=-v$. Adding these two identities gives $\sigma(\delta)=-\delta$.

Suppose $\delta=0$. Then $P+Q\sim2P_0$, so some nonconstant $h\in K(X)$ has divisor
\[
 (h)=P+Q-2P_0.
\]
Thus $1$ and $h$ are linearly independent elements of $\mathcal L(2P_0)$. Since $2\leq g$, Lemma~\ref{lem:low-degree} gives $h\in K(x)$. The divisor of poles of a function in $K(x)$ pulls back under the degree-two map $x$. Hence $h$ has no finite pole on $\mathbb P^1$, and its pole at infinity has order at most one. It follows that $h=ax+b$ with $a\ne0$. Every point in its zero divisor on $X$ has the same $x$-coordinate $-b/a$. This contradicts $x(P)\ne x(Q)$, so $\delta\ne0$.

If $u=v$, then $P\sim Q$. Since $P\ne Q$, a function with divisor $P-Q$ would give a degree-one map from $X$ to $\mathbb P^1$. Such a map identifies their function fields and would force $X$ to have genus zero. This contradicts $g\geq2$. Thus $u\ne v$.
\end{proof}

{Define
\begin{equation}
 \begin{split}
 \tau:J(K)&\longrightarrow J(K),\\
 \tau(z)&=\sigma(z)-\delta.
 \end{split}
 \label{eq:tau}
\end{equation}
\begingroup
For $z\in J(K)$, one has $\sigma^2(z)=z$, since $K=\F_{r^2}$. Lemma~\ref{lem:point-classes} gives $\sigma(\delta)=-\delta$. Therefore
\[
 \begin{aligned}
 \tau^2(z)&=\sigma\bigl(\sigma(z)-\delta\bigr)-\delta\\
 &=\sigma^2(z)-\sigma(\delta)-\delta\\
 &=z+\delta-\delta=z.
 \end{aligned}
\]
Thus $\tau$ is an involution.
\endgroup}

We next identify the required coset with the fixed points of an involution.

\begin{proposition}
\label{lem:frobenius-coset}
{For $\delta$ as in \eqref{eq:point-classes},} there is $b\in J(K)$ with $\sigma(b)-b=\delta$. {The fixed points of the involution $\tau$ in \eqref{eq:tau} are exactly $b+J(\F_r)$.}
\end{proposition}

\begin{proof}
The geometric points of the kernel of $\sigma-1$ form the finite group $J(\F_r)$. Since the differential of Frobenius is zero, the differential of $\sigma-1$ is $-\mathrm{id}$, so its kernel is finite and reduced. Thus $\sigma-1$ is an isogeny and is surjective on geometric points \cite[Part~I, Prop.~7.1]{Milne2008}. Choose $b\in J(\overline{\F}_r)$ satisfying
\begin{equation}
 \sigma(b)-b=\delta.
 \label{eq:Lang}
\end{equation}
Applying $\sigma$ and adding the resulting identity to \eqref{eq:Lang} gives $\sigma^2(b)-b=\sigma(\delta)+\delta=0$, so $b\in J(K)$.

Finally, \eqref{eq:Lang} shows that $\tau(z)=z$ exactly when $\sigma(z-b)=z-b$, that is, when $z-b\in J(\F_r)$. Hence the fixed points are $b+J(\F_r)$.
\end{proof}

\subsection{Intersection Formula and Fixed-Point Descent}
{
\begingroup
The next formula reduces an intersection involving degree $d$ to two translates involving degree $d-1$. Its genus-two case, $d=1$, appears in \cite[Prop.~4.5]{AreneKohelRitzenthaler2012}. In the proof of their Theorem~4.8, Frobenius permutes four theta translates, and a rational point would have to lie in their common intersection. We use the same general principle that a fixed point in exchanged sets lies in their intersection. The formula below allows this principle to reduce the degree by two in arbitrary genus.
\endgroup
}

\begin{lemma}
\label{lem:intersection}
Let $U,V\in X(K)$ be distinct, and set $u=[U-P_0]$, $v=[V-P_0]$. For $1\leq d\leq g-1$,
\begin{equation}
 \begin{split}
 W_d(K)\cap\bigl(W_d(K)+u-v\bigr)
 ={}&\bigl(u+W_{d-1}(K)\bigr)\\
 &\cup\bigl(-v+W_{d-1}(K)\bigr).
 \end{split}
 \label{eq:intersection}
\end{equation}
\end{lemma}

\begin{proof}
Let $z$ belong to the left side. By \eqref{eq:Wd}, there are effective $K$-rational divisors $A,B$ of degree $d$ such that
\[
 z=[A-dP_0],\qquad z-(u-v)=[B-dP_0].
\]
Subtracting these identities gives $A+V\sim B+U$. If $A$ contains $U$, the divisor $A-U$ is effective of degree $d-1$, and $z\in u+W_{d-1}(K)$.

Suppose instead that $A$ does not contain $U$. Since $U\ne V$, the divisor $A+V$ has coefficient zero at $U$, whereas $B+U$ has a positive coefficient there. These are distinct effective divisors in the same complete linear system. {\color{black}Their linear equivalence gives a nonconstant $h\in K(X)$ with $(h)=B+U-(A+V)$, so $h\in\mathcal L(A+V)$ and $\ell(A+V)\geq2$.} As $\deg(A+V)=d+1\leq g$, {Corollary~\ref{cor:low-degree-fiber}} gives an effective $K$-rational divisor $E$ of degree $d-1$ with $A+V\sim2P_0+E$. It follows that
\[
 z+v=[A+V-(d+1)P_0]=[E-(d-1)P_0],
\]
so $z\in-v+W_{d-1}(K)$. This proves one inclusion.

For the reverse inclusion, let $w=[E-(d-1)P_0]\in W_{d-1}(K)$. Then $u+w=[U+E-dP_0]\in W_d(K)$, and subtracting $u-v$ gives $v+w\in W_d(K)$. Similarly, \eqref{eq:fiber} gives $-v+w=[\iota(V)+E-dP_0]\in W_d(K)$. Subtracting $u-v$ now gives $-u+w=[\iota(U)+E-dP_0]\in W_d(K)$. Both sets on the right of \eqref{eq:intersection} therefore lie in the intersection.
\end{proof}

\begin{lemma}
\label{lem:fixed-point-descent}
For $1\leq d\leq g-1$, write $W_d(K)^\tau=\{z\in W_d(K):\tau(z)=z\}$. Then
\begin{equation}
 W_d(K)^\tau\subseteq
 (-u+W_{d-1}(K))\cap(-v+W_{d-1}(K)).
 \label{eq:fixed-subset}
\end{equation}
{Moreover, for $2\leq d\leq g-1$, one has $W_d(K)^\tau\subseteq W_{d-2}(K)$.}
\end{lemma}

\begin{proof}
We use the intersection formula twice.

Suppose $1\leq d\leq g-1$ and $z\in W_d(K)$ is fixed. Then $\sigma(z)=z+\delta$ belongs to $W_d(K)$, because this set is stable under $\sigma$. Therefore
\[
 z\in W_d(K)\cap\bigl(W_d(K)-\delta\bigr).
\]
Apply Lemma~\ref{lem:intersection} with $U=\iota(P)$ and $V=Q$. The distinct $x$-coordinates ensure $U\ne V$. Their divisor classes are $-u$ and $v$, so
\begin{equation}
 \begin{split}
 &W_d(K)\cap\bigl(W_d(K)-\delta\bigr)\\
 &\quad=(-u+W_{d-1}(K))\\
 &\qquad\quad\cup(-v+W_{d-1}(K)).
 \end{split}
 \label{eq:first-step}
\end{equation}
For every $w\in W_{d-1}(K)$, direct substitution in \eqref{eq:tau} gives
\[
 \begin{aligned}
 \tau(-u+w)&=-v+\sigma(w),\\
 \tau(-v+w)&=-u+\sigma(w).
 \end{aligned}
\]
Since $\sigma$ permutes $W_{d-1}(K)$, $\tau$ interchanges the two sets on the right of \eqref{eq:first-step}. A fixed point in their union must belong to their intersection, which proves \eqref{eq:fixed-subset}.

{\begingroup
Now assume $2\leq d\leq g-1$. A point $z$ in the intersection in \eqref{eq:fixed-subset} satisfies
\[
 z+u\in W_{d-1}(K)\cap
       \bigl(W_{d-1}(K)+u-v\bigr).
\]
Apply Lemma~\ref{lem:intersection} to $U=P$, $V=Q$, and degree $d-1$. This degree lies between $1$ and $g-2$. The resulting union is
\[
 (u+W_{d-2}(K))\cup(-v+W_{d-2}(K)).
\]
Subtracting $u$ gives
\endgroup
\begin{equation}
 \begin{split}
 &(-u+W_{d-1}(K))\cap(-v+W_{d-1}(K))\\
 &\qquad=W_{d-2}(K)\cup\bigl(W_{d-2}(K)-\delta\bigr).
 \end{split}
 \label{eq:second-step}
\end{equation}}

These two sets are again exchanged by $\tau$. Indeed, for $w\in W_{d-2}(K)$,
\[
 \begin{aligned}
 \tau(w)&=\sigma(w)-\delta,\\
 \tau(w-\delta)
 &=\sigma(w)-\sigma(\delta)-\delta\\
 &=\sigma(w).
 \end{aligned}
\]
Here we again use $\sigma(\delta)=-\delta$ and $\sigma(W_{d-2}(K))=W_{d-2}(K)$. Since $z$ is fixed and belongs to the union in \eqref{eq:second-step}, the same membership argument forces $z$ to belong to both sets. In particular, $z\in W_{d-2}(K)$, as required.
\end{proof}

\begin{proposition}
\label{prop:coset}
{Let $P,Q\in X(K)$ be $\iota\sigma$-fixed points with finite distinct $x$-coordinates. Then there is $b\in J(K)$ such that
\begin{equation}
 (b+J(\F_r))\cap W_d(K)=\varnothing
 \qquad(0\leq d\leq g-1).
 \label{eq:all-degrees}
\end{equation}}
\end{proposition}

\begin{proof}[Proof of Proposition~\ref{prop:coset}]
{Take $u,v,\delta$ from \eqref{eq:point-classes}.} Proposition~\ref{lem:frobenius-coset} gives $b\in J(K)$ such that the fixed points of $\tau$ are exactly $b+J(\F_r)$.

We prove by induction on $d$ that no point of $W_d(K)$ is fixed. For $d=0$, a fixed point would give $\tau(0)=0$, hence $\delta=0$, a contradiction. For $d=1$, Lemma~\ref{lem:fixed-point-descent} gives $W_1(K)^\tau\subseteq\{-u\}\cap\{-v\}=\varnothing$, since $u\ne v$. For $2\leq d\leq g-1$, a fixed $z\in W_d(K)$ would satisfy $z\in W_{d-2}(K)$ by the same lemma, contrary to the induction hypothesis. Therefore $(b+J(\F_r))\cap W_d(K)=\varnothing$ for every $0\leq d\leq g-1$, proving \eqref{eq:all-degrees}.
\end{proof}

}

\section{Proofs of the Main Results}
\label{sec:proofs}
\subsection{Proof of Theorem~\ref{thm:main}}
The geometric construction only requires two values of $x$ for which $f(x)$ is a nonsquare in the subfield. The Weil bound gives these values uniformly under the numerical hypothesis of Theorem~\ref{thm:main}. It also supplies the inequality needed to choose the effective representative.

\begin{lemma}
\label{lem:nonsquare-values}
Under the hypotheses of Theorem~\ref{thm:main}, one has $n<q=r^2$, and there are at least two distinct $a\in\F_r$ for which $f(a)$ is a nonsquare in $\F_r$.
\end{lemma}

\begin{proof}
The Weil bound \cite[Sec.~5.2]{Stichtenoth2009} states
\begin{equation}
 r+1-2g\sqrt r\leq n\leq r+1+2g\sqrt r.
 \label{eq:weil}
\end{equation}
Put $t=\sqrt r\geq2g+2$. Then $2gt\leq t^2-2t=r-2t$, so
\[
 n\leq2r+1-2t<2r<r^2,
\]
where $r\geq36$ has been used for the last inequality.

Let $\chi:\F_r\to\{-1,0,1\}$ be the quadratic character, equal to zero at zero, one at nonzero squares, and minus one at nonsquares. Write $B$ for the number of roots of $f$ in $\F_r$, $S$ for $\sum_{a\in\F_r}\chi(f(a))$, and $M$ for the number of nonsquare values. Counting the affine points, and then adding $P_0$, gives
\begin{equation}
 n=r+1+S,\qquad S=r-B-2M.
 \label{eq:character-count}
\end{equation}
Thus \eqref{eq:weil} already implies $|S|\leq2g\sqrt r$; no separate character-sum theorem is needed. Since $B\leq2g+1$,
\begin{equation}
 M\geq\frac{r-(2g+1)-2g\sqrt r}{2}
   =\frac{t(t-2g)-(2g+1)}2.
 \label{eq:nonsquares}
\end{equation}
Now $t(t-2g)\geq2t\geq4g+4$. The last expression is at least $(2g+3)/2>1$, proving the claim.
\end{proof}

{All parts of the construction are now available. We first find the coset and then apply the sufficient condition in Corollary~\ref{cor:coset-mds}.}

\begin{proof}[Proof of Theorem~\ref{thm:main}]
Let $K=\F_{r^2}$. Lemma~\ref{lem:nonsquare-values} gives distinct $a_1,a_2\in\F_r$ with $f(a_i)$ nonsquare. Each $f(a_i)$ has a square root $y_i$ in $K$, since its quadratic polynomial is irreducible over $\F_r$ and splits over its quadratic extension. The nontrivial automorphism of $K/\F_r$ interchanges the roots, so $y_i^r=-y_i$. Hence the points $P=(a_1,y_1)$ and $Q=(a_2,y_2)$ satisfy \eqref{eq:points}. {Proposition~\ref{prop:coset} gives $b\in J(K)$ satisfying \eqref{eq:all-degrees}; taking $d=g-1$ gives \eqref{eq:coset}.}

{Lemma~\ref{lem:nonsquare-values} also gives $n<q$. By \eqref{eq:dimension}, $k\geq g+1$ and $k+g-1\leq n-1<n$. Corollary~\ref{cor:coset-mds} therefore applies with this $b$ and $k$. It gives an effective $K$-rational divisor $G$ of degree $k+g-1$, with support disjoint from $D$, such that $C_{\mathcal L}(D,G)$ is an $[n,k,n-k+1]_K$ MDS code. Since $K=\F_q$, this proves the stated parameters and support condition.}
\end{proof}

\subsection{Non-equivalence to GRS Codes}
{This subsection concerns only code equivalence under the hypothesis $\min\{k,n-k\}\geq3g+1$. It is independent of the MDS construction and is not used to establish the MDS property.}

{
\begin{definition}
\label{def:grs}
Two length-$n$ linear codes over $K$ are monomially equivalent if one is obtained from the other by a permutation of coordinates and multiplication of each coordinate by a nonzero scalar. For distinct $a_1,\ldots,a_n\in K$ and $v_1,\ldots,v_n\in K^\times$, a generalized Reed--Solomon (GRS) code of dimension $k$ is
\[
 \{(v_1h(a_1),\ldots,v_nh(a_n)):
          h\in K[x],\ \deg h<k\}.
\]
\end{definition}

\begin{definition}
\label{def:schur-square}
For $c,c'\in K^n$, put $c*c'=(c_1c'_1,\ldots,c_nc'_n)$. The Schur square of a linear code $C$ is
\[
 C^{*2}=\operatorname{span}_K\{c*c':c,c'\in C\}.
\]
\end{definition}
}
Its dimension is invariant under monomial equivalence. Indeed, a permutation acts on the square by the same permutation, and coordinate multipliers $v_i$ act by the nonzero multipliers $v_i^2$.

\begingroup
Schur-square dimensions already distinguish AG codes in \cite[Thm.~4.1]{Chen2024}; see also \cite[Lem.~8]{Zhu2026}. We use this established method with general defining divisors. Shortening will ensure injectivity after squaring, including at rate $1/2$. We first state the standard multiplication theorem needed for those divisors.
\endgroup

\begin{lemma}
\label{lem:multiplication}
Let $X/K$ be a smooth projective geometrically irreducible curve of genus $g$, and let $H$ be a $K$-rational divisor of degree at least $2g+1$. Then the map
\[
 \mathcal L(H)\otimes_K\mathcal L(H)
       \longrightarrow\mathcal L(2H),\qquad
       h_1\otimes h_2\longmapsto h_1h_2
\]
is surjective.
\end{lemma}

\begin{proof}
Over an algebraic closure $\overline K$, this is the standard multiplication theorem for line bundles of degree at least $2g+1$ \cite[Thm.~6]{Mumford1970}. Global sections commute with field extension for these line bundles. Hence tensoring the displayed map with $\overline K$ gives that surjective map. Its cokernel therefore becomes zero after tensoring with $\overline K$. Faithful flatness of the field extension implies that the cokernel over $K$ was already zero.
\end{proof}

The next proposition distinguishes the resulting codes from GRS codes at every fixed limiting rate, including rate $1/2$. The shortening step ensures that evaluation remains injective after squaring.

\begin{proposition}
\label{prop:non-grs}
Let $X/K$ have genus $g\geq2$, and let $D$ be the sum of $n$ distinct $K$-rational points. Let $G$ be a $K$-rational divisor with support disjoint from $D$, and suppose that $C=C_{\mathcal L}(D,G)$ is an MDS code of dimension $k$ with
\[
 \deg G=k+g-1<n,\qquad
 \min\{k,n-k\}\geq3g+1.
\]
Then $C$ is not monomially equivalent to a GRS code.
\end{proposition}

\begin{proof}
The dual of a GRS code is a GRS code, and taking duals preserves monomial equivalence. The duality theorem for AG codes \cite[Thm.~2.2.8 and Prop.~2.2.10]{Stichtenoth2009} expresses $C^\perp$, up to nonzero coordinate multipliers, as
\[
 C_{\mathcal L}(D,H_0),\qquad
 \deg H_0=n-k+g-1,
\]
where $H_0$ is defined over $K$ and has support disjoint from $D$. One obtains $H_0=D-G+(\eta)$ by choosing a differential $\eta$ with a simple pole at every evaluation point \cite[Lem.~2.2.9]{Stichtenoth2009}; its residues give the coordinate multipliers. Since $k\geq3g+1$, this degree is less than $n$. Replacing $C$ by that evaluation code if necessary, we may assume $k\leq n/2$.

Put $t=2g-1$ and choose any sum $T$ of $t$ distinct points in $D$. The code shortened on these coordinates consists of the codewords that are zero there, followed by deletion of those coordinates. Since $G$ avoids $D$, it is exactly
\[
 C'=C_{\mathcal L}(D-T,G-T).
\]
Its dimension is $k'=k-t$. This follows either from the MDS property or from Riemann--Roch, since, on putting $H=G-T$, we have
\[
 \deg H=k-g\geq2g+1.
\]
Write $n'=n-t$. The assumption $2k\leq n$ gives
\[
 2\deg H=2k-2g\leq n-2g<n-2g+1=n'.
\]
Lemma~\ref{lem:multiplication} therefore gives
\[
 (C')^{*2}=C_{\mathcal L}(D-T,2H),
\]
and the last degree inequality makes the evaluation map on $\mathcal L(2H)$ injective. Riemann--Roch yields
\begin{equation}
 \dim_K(C')^{*2}
     =2\deg H+1-g
     =2k'+g-1.
 \label{eq:shortened-square}
\end{equation}

Shortening a GRS code on $t<k$ coordinates gives a GRS code of length $n'$ and dimension $k'$. Indeed, the polynomial being evaluated must be divisible by the product of the $t$ corresponding linear factors. Factoring out that product gives the new nonzero coordinate multipliers. Products of polynomials of degree less than $k'$ span all polynomials of degree at most $2k'-2$. Thus the Schur square of this shortened GRS code has dimension
\[
 \min\{n',2k'-1\}=2k'-1,
\]
because $2k'-1=2k-4g+1<n'$. This differs from \eqref{eq:shortened-square} by $g>0$. Monomial equivalence would preserve the dimension of the Schur square after the corresponding shortening. This contradiction proves the proposition.
\end{proof}

In particular, for fixed $g$ and $k/n\to\rho\in(0,1)$, the condition $\min\{k,n-k\}\geq3g+1$ holds for all sufficiently large $n$. This proves the non-GRS assertions in the asymptotic corollaries without excluding half-rate codes.

\subsection{Proof of Corollary~\ref{cor:rate}}
\begin{proof}[Proof of Corollary~\ref{cor:rate}]
{We may assume that $r=\sqrt q\geq(2g+2)^2$.} Since $r>2g+1$, choose distinct $c_1,\ldots,c_{2g+1}\in\F_r$ and put
\[
 f(x)=\prod_{i=1}^{2g+1}(x-c_i).
\]
This is square-free of degree $2g+1$. The finite branch points and $P_0$ give $2g+2$ rational points fixed by $\iota$. All other rational points occur in pairs. Therefore $n=|X(\F_r)|$ is even.

With $t=\sqrt r\geq2g+2$, the lower Weil bound gives
\[
 n\geq t(t-2g)+1\geq2t+1\geq4g+5>2g+2.
\]
{Substituting $r=\sqrt q$ into \eqref{eq:weil} gives \eqref{eq:length}, and dividing it by $\sqrt q$ proves $n/\sqrt q\to1$ for fixed $g$.}

For a fixed $0<\rho<1$, both $\lfloor\rho n\rfloor$ and $n-\lfloor\rho n\rfloor$ tend to infinity linearly with $n$. {
{Hence they are at least $3g+1$ for all sufficiently large $q$. Theorem~\ref{thm:main} gives the MDS parameters, and Proposition~\ref{prop:non-grs} gives non-equivalence to GRS codes.}
} {Finally, $k=\lfloor\rho n\rfloor$ and \eqref{eq:length} prove \eqref{eq:rates}.}
\end{proof}

{\noindent\textbf{Remark.} Put $c_\rho=\min\{\rho,1-\rho\}$ and
\[
 R_{g,\rho}=\max\left\{2g+2,\,
 \frac{(3g+1)c_\rho^{-1}-1}{2}\right\}.
\]
If $q\geq R_{g,\rho}^4$, then the lower Weil bound gives $n\geq2q^{1/4}+1\geq(3g+1)c_\rho^{-1}$. Hence $k=\lfloor\rho n\rfloor$ satisfies $\min\{k,n-k\}\geq3g+1$. This is an explicit sufficient bound for both the MDS and the non-GRS assertions in Corollary~\ref{cor:rate}.\par}
{
\begin{corollary}
\label{cor:half-rate}
Fix $g\geq2$. For every odd square $q$ with $\sqrt q\geq(2g+2)^2$, there are genus-$g$ MDS algebraic geometry codes with parameters $[n,n/2,n/2+1]_q$, where $n$ is even and satisfies \eqref{eq:length}. If $n\geq6g+2$, these codes are not monomially equivalent to GRS codes.
\end{corollary}
This follows from the parity and dimension checks in the proof of Corollary~\ref{cor:rate} and Theorem~\ref{thm:main}. The non-equivalence assertion follows from $k=n-k=n/2\geq3g+1$.
}

\subsection{Proof of Corollary~\ref{cor:exact}}

{A curve of genus $g$ over $\F_{s^2}$ is called maximal if it has $s^2+1+2gs$ rational points.}

{
\begin{lemma}
\label{lem:maximal-point-count}
Fix $g\geq2$ and a prime $p\equiv-1\pmod{2(2g+1)}$. The curve $X$ in \eqref{eq:maximal} is maximal over $\F_{p^2}$. Moreover, for every positive odd integer $m$ and $r=p^{2m}$,
\begin{equation}
 |X(\F_r)|=r+1+2gp^m.
 \label{eq:maximal-count}
\end{equation}
\end{lemma}

\begin{proof}
Put $d=2g+1$. The congruence on $p$ gives $2d\mid p+1$. In particular, $p$ is odd and does not divide $d$, so $1-x^d$ is square-free. Over $\F_{p^2}$, take the Hermitian curve with affine equation $u^{p+1}+v^{p+1}=1$. The formulas
\begin{equation}
 x=u^{(p+1)/d},\qquad y=v^{(p+1)/2}
 \label{eq:hermitian-map}
\end{equation}
satisfy $y^2=1-x^d$. {They define a nonconstant rational map to $X$. This map extends to a finite surjective morphism between the smooth projective models over $\F_{p^2}$.} Since the Hermitian curve is maximal, the covered curve $X$ is maximal over that field \cite[Sec.~1]{GarciaStichtenothXing2000}.

The Frobenius eigenvalues $\alpha_1,\ldots,\alpha_{2g}$ of $X$ over $\F_{p^2}$ have modulus $p$ and sum $-2gp$ by maximality. Equality in the triangle inequality forces every $\alpha_i=-p$. Since $m$ is odd,
\[
 \begin{aligned}
 |X(\F_r)|&=r+1-\sum_{i=1}^{2g}\alpha_i^m\\
 &=r+1-2g(-p)^m\\
 &=r+1+2gp^m,
 \end{aligned}
\]
which is \eqref{eq:maximal-count}.
\end{proof}
}

\begin{proof}[Proof of Corollary~\ref{cor:exact}]
{Put $d=2g+1$ and $r=p^{2m}$, so that $q=r^2$. The congruence gives $p\geq4g+1$ and $p\nmid d$. Thus $1-x^d$ is square-free and $r\geq(2g+2)^2$. Lemma~\ref{lem:maximal-point-count} gives the point count \eqref{eq:maximal-count}.}

{Formula \eqref{eq:maximal-count} gives $n\longrightarrow\infty$ through odd $m$. For fixed $0<\rho<1$, both $k=\lfloor\rho n\rfloor$ and $n-k$ are therefore at least $3g+1$ for all sufficiently large odd $m$, so \eqref{eq:dimension} holds. {Theorem~\ref{thm:main} gives \eqref{eq:rho-parameters}, and Proposition~\ref{prop:non-grs} gives non-equivalence to GRS codes.} Finally, $r=\sqrt q$ and $p^m=q^{1/4}$ turn \eqref{eq:maximal-count} into \eqref{eq:exact}.}
\end{proof}

\subsection{Examples and the Scope of the Construction}
We check the field sizes, point counts, and dimension conditions in three examples. These checks illustrate the general theorem, the half-rate consequence, and the exact family, respectively.

\begin{example}
  Take $g=2$, $r=41$, and $X:y^2=1-x^5$. Then $q=1681$ and $r\geq36$. The polynomial is square-free. In the notation of \eqref{eq:character-count}, direct calculation gives $B=5$ and $M=10$, whence
\[
 S=41-5-2\cdot10=16,\qquad n=41+1+16=58.
\]
For example, $f(6)=f(14)=15$ in $\F_{41}$, and $15$ is a nonsquare. Thus the two points required in the proof can be chosen above $6$ and $14$. The range \eqref{eq:dimension} is {
$3\leq k\leq56$.
} Taking $k=20$ gives an $[58,20,39]_{1681}$ code, with a defining divisor of degree $21$.
\end{example}

\begin{example}
  Take $g=2$, $r=41$, and
\[
 f(x)=x(x-1)(x-2)(x-3)(x-4).
\]
Its five distinct roots and $P_0$ are rational. Here $B=5$ and $M=12$, so $S=12$ and $n=54$. {Corollary~\ref{cor:half-rate} with $k=27$ gives an $[54,27,28]_{1681}$ code, with a defining divisor of degree $28$.}
\end{example}

\begin{example}
  Take $g=3$, $p=13$, and $m=1$ in Corollary~\ref{cor:exact}. The congruence $13\equiv-1\pmod{14}$ holds. We have $r=169\geq64$, $q=28561$, and
\[
 n=169+1+6\cdot13=248.
\]
{Taking $\rho=1/2$ gives $k=124$, and $4\leq124\leq245$, so the dimension condition holds already at $m=1$. Thus we obtain an $[248,124,125]_{28561}$ code, with a defining divisor of degree $126$.}
\end{example}
{
All three examples also satisfy the hypothesis of Proposition~\ref{prop:non-grs}. Table~\ref{tab:shortened-examples} gives the exact Schur-square dimensions after shortening on $2g-1$ coordinates. The strict difference proves non-equivalence to GRS codes; it is not a numerical assumption about a chosen generator matrix.

\begin{table}[t]

\caption{Schur squares after shortening the three examples.}
\label{tab:shortened-examples}
\centering
\footnotesize
\setlength{\tabcolsep}{3pt}
\begin{tabular}{@{}cccc@{}}
\toprule
Original $[n,k]$ & Shortened $[n',k']$ &
$\dim(C')^{*2}$ & GRS square\\
\midrule
$[58,20]$ & $[55,17]$ & $35$ & $33$\\
$[54,27]$ & $[51,24]$ & $49$ & $47$\\
$[248,124]$ & $[243,119]$ & $240$ & $237$\\
\bottomrule
\end{tabular}
\end{table}
}

\begingroup
The point counts above verify the numerical inputs. The MDS conclusion follows from the geometric proof. We now specify a defining class, an effective divisor, and a generator matrix for the first example. This also separates a complete code construction from a check of its numerical parameters.
\endgroup

\begingroup
\subsection{\texorpdfstring{A Fully Specified Code over $\F_{1681}$}{A Fully Specified Code over F1681}}
Let $K=\F_{41}(\theta)$ with $\theta^2=3$. Since $3$ is a nonsquare in $\F_{41}$, this is a quadratic field, and $\sigma(\theta)=-\theta$. Keep $X:y^2=f(x)=1-x^5$ and the evaluation divisor $D$ consisting of all $58$ points in $X(\F_{41})$. Set
\[
 P=(6,13\theta),\qquad Q=(14,13\theta).
\]
Here $(13\theta)^2=15=f(6)=f(14)$, so these points satisfy \eqref{eq:points}. Define
\[
 \begin{aligned}
 u_R(x)&=x^2+(35+21\theta)x+23,\\
 v_R(x)&=(17+5\theta)x+33+40\theta.
 \end{aligned}
\]
The congruence $v_R^2\equiv f\pmod{u_R}$ and the square-freeness of $u_R$ define an effective divisor $R$ of degree two by $u_R(x)=0$, $y=v_R(x)$. Its Mumford class is $b=[R-2P_0]$.

Here is an exact certificate for \eqref{eq:Lang}. Put
\[
 \begin{aligned}
 u_\delta(x)&=(x-6)(x-14),\\
 A(x)&=x^3+x^2+33x+9,\qquad C=37\theta.
 \end{aligned}
\]
Polynomial multiplication and division over $K$ give
\[
 \begin{aligned}
 A-Cv_R&\equiv0\pmod{u_R},\\
 A+C\sigma(v_R)&\equiv0\pmod{\sigma(u_R)},\\
 A^2-C^2f&=u_\delta u_R\sigma(u_R),
 \end{aligned}
\]
where $C^2=7$. The three factors on the last right-hand side are square-free and pairwise coprime. At $P,Q$, the numerator $A+Cy$ is nonzero, and at $\iota(P),\iota(Q)$ it vanishes. Its other zeros are precisely $\iota(R)+\sigma(R)$. Its pole at $P_0$ has order six. Consequently,
\[
 \varphi=\frac{A+Cy}{u_\delta},\qquad
 (\varphi)=\sigma(R)+\iota(R)-P-Q-2P_0.
\]
Since $R+\iota(R)\sim4P_0$, this proves
\[
 \sigma(b)-b=[P+Q-2P_0]=\delta.
\]
Proposition~\ref{prop:coset} therefore applies to this specified class $b$.

To choose the effective representative for $k=20$, put
\[
 \psi=\frac{y+v_R(x)}{u_R(x)},\qquad
 G_0=R+19P_0,\qquad h=x^9\psi+1.
\]
The poles of $\psi$ are the two points in $R$ and a pole of order one at $P_0$. Thus $h\in\mathcal L(G_0)$ and its pole at $P_0$ has order $19$. We check that $h$ is nonzero at every finite evaluation point. For $a\in\F_{41}$, $u_R(a)\ne0$: its $\theta$-coefficient is $21a$, and $u_R(0)=23$. If $h(a,c)=0$ with $a,c\in\F_{41}$, the $\theta$-coefficient of
\[
 a^9\bigl(c+v_R(a)\bigr)+u_R(a)=0
\]
would give $5a^{10}+40a^9+21a=0$. The exact identity
\[
 \gcd(x^{41}-x,\,5x^{10}+40x^9+21x)=x(x-15)
\]
leaves only $a=0$ or $a=15$. At $a=0$ the numerator is $23$, whereas $f(15)=28$ is a nonsquare in $\F_{41}$. Neither case gives a zero at an evaluation point. Hence
\[
 G=G_0+(h)
\]
is effective of degree $21$ and avoids $D$, including $P_0$. Moreover,
\[
 [G-21P_0]=[G_0-21P_0]=[R-2P_0]=b.
\]
The code $C_{\mathcal L}(D,G)$ is therefore $[58,20,39]_{1681}$ by Corollary~\ref{cor:coset-mds}.

For completeness, a basis of $\mathcal L(G_0)$ is
\[
 1,x,\ldots,x^9,\quad
 \psi,x\psi,\ldots,x^9\psi.
\]
The pole orders at $P_0$ are the distinct integers $0,1,\ldots,19$, so these functions are independent. Riemann--Roch gives $\ell(G_0)=20$, proving that the list is a basis of the full space. Dividing the list by $h$ gives a basis of $\mathcal L(G)$.

Order the finite points $R_j=(a_j,c_j)$ of $X(\F_{41})$ lexicographically using representatives $0,\ldots,40$, and put $R_{58}=P_0$. An explicit $20\times58$ generator matrix $M$ is specified by
\[
 \begin{aligned}
 M_{i+1,j}&=\frac{a_j^i}{h(R_j)},\\
 M_{i+11,j}&=\frac{a_j^i\psi(R_j)}{h(R_j)}
 \end{aligned}
 \qquad(0\leq i\leq9,\ 1\leq j\leq57).
\]
The last column has entry $1$ in row $20$ and zero elsewhere. This follows by taking values at $P_0$: $h/(x^9\psi)$ tends to $1$, and all other basis elements have smaller pole order. Exact row reduction gives rank $20$. Shortening on the first three columns gives rank $17$ and Schur-square rank $35$, agreeing with Table~\ref{tab:shortened-examples}. These computations check the specified matrix; its MDS property follows from the divisor certificate and the coset proof.
\endgroup

\section{Future Work}
\label{sec:future}
{Extending the construction beyond hyperelliptic curves requires a new proof of the coset condition \eqref{eq:coset}: Lemma~\ref{lem:intersection} uses the degree-two map essentially.}

{
\begingroup
The specified example shows how a Lang solution and a section give a generator matrix. A general algorithm must perform these steps uniformly: solve \eqref{eq:Lang}, represent its solution by a divisor, choose a section as in Lemma~\ref{lem:effective-representative}, and compute a basis of $\mathcal L(G)$. A complexity theorem requires analyzing each step.
\endgroup

Proposition~\ref{prop:non-grs} proves non-equivalence to GRS codes when $\min\{k,n-k\}\geq3g+1$. Further equivalence questions concern different choices of the defining coset and comparisons with other non-GRS constructions, including \cite{Zhu2026,Marques2026}. Determining the maximal attainable length and improving the leading constant of the present construction remain separate problems. The existence lower bound proved here supplies no matching upper bound.

Finally, the present argument does not give corresponding families in characteristic two, over nonsquare alphabet fields, or with a one-point defining divisor. These are questions for extending this construction, rather than claims that MDS constructions in those settings are unknown.
}

\section*{Acknowledgment}
The authors acknowledge the assistance of ChatGPT 6 (\mbox{OpenAI}) in mathematical discussions and {polishing the manuscript and correcting its grammar}. {
Chuangqiang Hu was supported by the National Natural Science Foundation of China (No.~12441107).
}

\end{document}